\documentclass[%
 reprint,
superscriptaddress,
 amsmath,amssymb,
 aps,
pra,
]{revtex4-2}

\setcitestyle{super}
\usepackage{graphicx}
\usepackage{dcolumn}
\usepackage{bm}

\usepackage[separate-uncertainty=true]{siunitx}

\newcommand{\Sto}{$\vert\mathrm{S21}\vert$ }
\newcommand{\inlinecite}[1]{[\citenum{xuFloquetCavityElectromagnonics2020}]}%

\begin{document}

\preprint{APS/123-QED}

\title{Multifrequency Floquet Engineering of Magnon Polaritons}

\author{L.~Hackner}
\affiliation{Department of Physics, University of Otago, Dunedin, New Zealand}
\affiliation{Dodd-Walls Centre for Photonic and Quantum Technologies, Dunedin, New Zealand}

\author{A.~R.~Myatt}
\affiliation{Department of Physics, University of Otago, Dunedin, New Zealand}
\affiliation{Dodd-Walls Centre for Photonic and Quantum Technologies, Dunedin, New Zealand}

\author{W.~Wustmann}
\affiliation{Department of Physics, University of Otago, Dunedin, New Zealand}
\affiliation{Dodd-Walls Centre for Photonic and Quantum Technologies, Dunedin, New Zealand}

\author{N.~J.~Lambert}
\email[Contact author: ]{nicholas.lambert@otago.ac.nz}
\affiliation{Department of Physics, University of Otago, Dunedin, New Zealand}
\affiliation{Dodd-Walls Centre for Photonic and Quantum Technologies, Dunedin, New Zealand}

\date{\today}

\begin{abstract}
Floquet engineering of cavity magnon-polaritons by periodically modulating the magnon frequency has recently attracted much interest as a way to manipulate the energy spectrum of magnon-photon hybrid systems. However, modulating the frequency of magnons by a time-varying bias magnetic field can be challenging.  We demonstrate cavity magnon-polariton Floquet engineering by modulating the microwave cavity frequency, allowing large modulation depth and bandwidth. We apply commensurate two-frequency Floquet modulations with the higher frequency at twice and three times the lower frequency, and demonstrate that the resulting spectrum depends on the relative amplitude and phase of the two drive tones. In comparison with single-frequency Floquet modulations, the spectrum has qualitatively different features; in particular, new anticrossings appear between previously uncoupled sidebands. Our platform offers an alternative way to manipulate Floquet quasi-energy levels in hybrid systems.

\end{abstract}

\maketitle


Cavity electromagnonic systems are a promising emerging platform for quantum technologies~\cite{lachance-quirionHybridQuantumSystems2019, zarerameshtiCavityMagnonics2022}. In these hybrid devices, a magnetically ordered material exhibiting well-defined magnetostatic resonances is embedded in an electromagnetic cavity or resonator such that the supported magnons can exchange energy with the photons. If the mode overlap between the two excitations is sufficiently large, the coupling rate between the two exceeds the dissipation rates in the system, and the strong coupling regime is reached~\cite{goryachevHighCooperativityCavityQED2014, tabuchiHybridizingFerromagneticMagnons2014, zhangStronglyCoupledMagnons2014,lambertIdentificationSpinWave2015}. Judicious selection of cavity or magnon host material can allow ultrastrong coupling to be observed~\cite{bourhillUltrahighCooperativityInteractions2016, evertsUltrastrongCouplingMicrowave2020a}. The resulting hybrid excitations are termed magnon-polaritons, and demonstrations of hybrid devices based on magnons have included strong coupling between magnons and qubits\cite{tabuchiCoherentCouplingFerromagnetic2015}, strong coupling between spatially separated magnon populations\cite{lambertCavitymediatedCoherentCoupling2016a}, and non-reciprocal devices\cite{wangNonreciprocityUnidirectionalInvisibility2019, zhangBroadbandNonreciprocityEnabled2020}. They have been proposed as a candidate platform for a variety of quantum technologies including coherent transducers~\cite{hisatomi, osada, PhysRevLett.117.123605, zhuWaveguideCavityOptomagnonics2020} and quantum memories~\cite{fergusonGenerationLightMultimode2016a, tanjiHeraldedSingleMagnonQuantum2009, zhangMagnonDarkModes2015}.

There has been significant recent interest in temporal periodic modulation of the parameters of cavity electromagnonic systems, resulting in changes in their dynamics. This powerful technique, termed Floquet engineering, introduces controllable quasienergy levels into the system's spectrum. Despite the significant experimental difficulties associated with rapid tuning of the magnon modes, there have been a number of demonstrations of Floquet physics in magnonic systems~\cite{xuFloquetCavityElectromagnonics2020}.
The technique has been applied to a range of problems, such as the engineering of magnon dark modes~\cite{pishehvarOndemandMagnonResonance2025}, controllable magnon-magnon coupling\cite{heidarpourFloquetdrivenControlIndirect2025}, and the generation of entangled states~\cite{zhuFloquetengineeringMagnonicNOON2023, qiFloquetGenerationMagnonic2023}. The new energy bands created are also of more fundamental interest; they have non-trivial topological properties~\cite{chenFloquetNogoTheorem2025, minguzziTopologicalPumpingFloquetBloch2022, olinTopologicalPhaseTransition2023}, and have been demonstrated to exhibit novel phases~\cite{olinTopologicalPhaseTransition2023}. Recently, the difficulty of achieving large frequency modulations was tackled by using a resonantly-enhanced drive field~\cite{pishehvarResonanceenhancedFloquetCavity2025}.

However, there remain significant experimental challenges associated with rapid tuning of the magnon modes. A time-varying magnetic field of a relatively large amplitude must be applied over the volume of the magnetic material, preferably via a coil or antenna with a low inductance and a flat frequency response over a broad range. Furthermore, the field should be as uniform as possible to avoid inhomogeneous broadening, of and coupling between, different magnetostatic modes~\cite{schlomannInhomogeneousBroadeningFerromagnetic1969, PhysRevB.97.214423}. Here, we avoid the difficulties of modulating the magnon frequency by instead employing a rapidly tunable microwave resonator. Whilst the symmetry between the magnon and photon modes in the equations describing the system means that the underlying physics remains the same, our approach allows for large amplitudes and frequencies of modulation, and furthermore arbitrary and precisely controlled modulation waveforms can be applied. We use this approach to study dual-frequency Floquet driving of the hybrid system, focusing primarily on the case where one frequency is twice the other, and the case where one frequency is three times the other.

\begin{figure}
\includegraphics{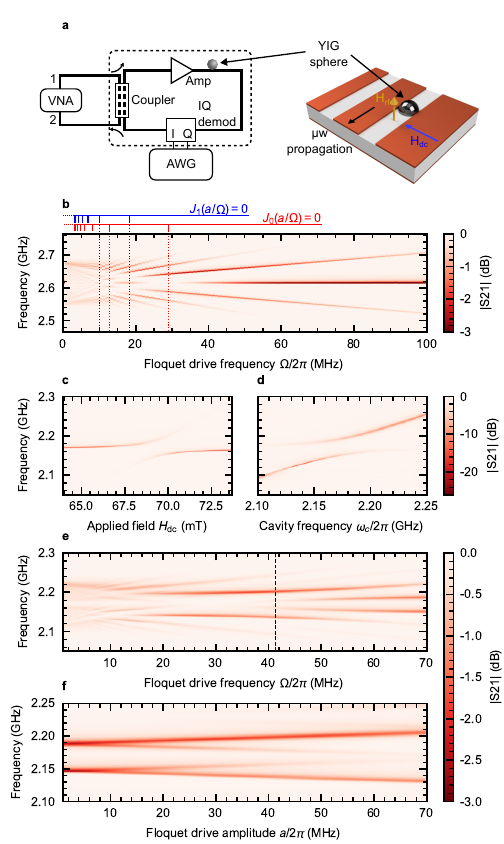}
\caption{\label{fig:ExperimentConfiguration} \textbf{Experimental setup and device characterisation.} (a) Schematic the hybrid magnonic device comprising a microwave ring resonator with gain (Amp) and phase control (IQ), and an embedded YIG sphere (right). A vector network analyser (VNA) measures the transmission (S21) of a feedline, and an arbitrary waveform generator (AWG) allows the frequency of the cavity modes to be modulated. 
(b) \Sto of uncoupled cavity mode spectrum under single tone frequency modulation as function of drive frequency $\Omega/2\pi$, with $\omega_c/2\pi=\qty{2.615}{\giga\hertz}$, showing sidebands with minima coincident with Bessel function zeros (top axis, dotted lines.) (c) \Sto as a function of applied magnetic field, with  $\omega_c/2\pi=\qty{2.17}{\giga\hertz}$, showing a magnon-photon anticrossing of size $2g/2\pi=\qty{41.6}{\mega\hertz}$. (d) \Sto as a function of cavity frequency $\omega_c$, with magnetic field $H_\textrm{dc}$ fixed at \qty{69}{\milli\tesla}, again showing an anticrossing. (e) \Sto as a function of Floquet drive frequency $\Omega/2\pi$, with $\omega_c/2\pi = \omega_m/2\pi= \qty{2.17}{\giga\hertz}$ and $a/2\pi=\qty{50}{\mega\hertz}$. (f) \Sto of feedline as a function of probe frequency and Floquet drive amplitude, with $\omega_c/2\pi = \omega_m/2\pi= \qty{2.17}{\giga\hertz}$, and $\Omega/2\pi=\qty{41.4}{\mega\hertz} = 2g/2\pi$, corresponding to the dashed line in panel (e).}
\end{figure}

Our device (Fig.~\ref{fig:ExperimentConfiguration}(a)) comprises a loop of coplanar waveguide on a low-loss printed circuit board (PCB) (Rogers 4003C) with an embedded amplifier (Mini-Circuits YSF-322+) and an in-phase/quadrature (IQ) demodulator (Marki Microwave MMIQ-0205HSM-2). We also add a backwards-wave stripline directional coupler (-16dB coupling, Mini-Circuits SCBD-16-63HP+) to the loop to provide coupling to a feedline. Resonant modes occur when the propagating microwave field undergoes a $2n\pi$ (where $n$ is an integer) phase shift during one round trip of the loop. The phase shift has a controllable component due to the IQ demodulator; applying carefully calibrated voltages to the I and Q ports of the demodulator allows high bandwidth control of the frequency of the modes. The amplitude of the propagating wave is also dependent on the I and Q voltages, thus allowing control over the total energy loss per round trip and hence the internal loss rate of the cavity mode. (Here we keep the internal loss of the mode constant throughout.)

A polished yttrium iron garnet (YIG) sphere\footnote{{www.ferrisphere.com}} of diameter \qty{1}{\milli\metre} is inserted between the centre conductor and the ground plane of the waveguide loop such that the magnetic field component of the microwave field through the sphere is normal to the PCB. YIG is a common material for hybrid magnonic devices due to its high spin density and low loss, helping to reach the strong coupling regime between magnons and photons; a spherical geometry gives a well understood family of modes~\cite{PhysRev.105.390, PhysRev.114.739}.

To calibrate the I and Q voltages required for a particular set of cavity parameters, the magnon population is detuned from the cavity by applying a magnetic field such that the magnon frequency $\omega_m/2\pi$ is above $\qty{4}{\giga\hertz}$. We apply static voltages from an arbitrary waveform generator (AWG) and measure the transmission spectrum of the feedline from port 1 to port 2 (S21) using a vector network analyser (VNA), as in Fig.~\ref{fig:ExperimentConfiguration}(a). We observe an absorption dip due to the cavity resonance, and determine the centre frequency and linewidth from a Lorentzian fit. By iteratively tuning the applied voltages until the centre frequency and linewidth of the resonance matches a particular target frequency and loss rate, we find the trajectory in (I,Q) voltage space corresponding to frequency modulation with constant internal loss rate.

We then use the AWG to apply waveforms corresponding to a single tone Floquet drive such that the time-varying angular frequency of the microwave mode is given by $\omega(t) = \omega_c + a \sin (\Omega t)$, where $\Omega$ and $a$ are the Floquet drive frequency and drive amplitude, respectively. In Fig.~\ref{fig:ExperimentConfiguration}(b) we show \Sto for the modulated cavity (with the magnon frequency still far detuned from the cavity frequency) as a function of $\Omega/2\pi$, with $\omega_c/2\pi = \qty{2.615}{\giga\hertz}$ and $a/2\pi=\qty{70}{MHz}$. Sidebands appear at $\omega_c\pm n\Omega$, with integer $n$. Furthermore, we find that the $n$th sideband disappears at the zeros of the Bessel function $J_n(a/\Omega)$ (top axis) as expected~\cite{xuFloquetCavityElectromagnonics2020}, confirming that our I and Q voltage calibration still applies with higher frequency drives and is not bandwidth limited in any way.

When the applied magnetic field is such that the magnon frequency is similar to the photon frequency, the excitations of the system become hybridized magnon-polaritons. We probe the spectrum of the hybrid system by measuring S21. We begin by tuning the cavity frequency to  \qty{2.17}{GHz}. In all the following experiments we fix the internal loss rate of the cavity to be $2\pi\times\qty{2}{\mega\hertz}$, and we determine the external loss rate (set by the coupling of the directional coupler) to be $2\pi\times\qty{1.6(0.05)}{\mega\hertz}$. In order to tune the magnon mode frequency through the cavity frequency, we sweep an in-plane magnetic field $H_{\text{dc}}$ from \qty{64}{\milli\tesla} to \qty{73.7}{\milli\tesla}. We observe an anticrossing between the spatially uniform (Kittel) magnon mode supported by the YIG and photons in the cavity (Fig.~\ref{fig:ExperimentConfiguration}(c)), demonstrating that the two are in the strong coupling regime. By fitting the eigenmodes of two coupled oscillators to the observed anticrossing, we determine a coupling rate of $g/2\pi = \qty{20.7}{\mega\hertz}$. We next fix the applied field at \qty{69}{\milli\tesla}, resulting in a bare magnon frequency of \qty{2.17}{\giga\hertz}, and use the tunability of our microwave cavity to sweep the photon frequency from \qty{2.1}{\giga\hertz} to \qty{2.25}{\giga\hertz}. We again see an anticrossing between the modes (Fig.~\ref{fig:ExperimentConfiguration}(d)), and find a minimum mode separation of $2g/2\pi = 2\times \qty{20.5}{\mega\hertz}$, in reasonable agreement with the previous measurement.


We now set the cavity mode centre frequency to $\omega_c/2\pi = \qty{2.17}{\giga\hertz}$, and apply a magnetic field of \qty{69}{\milli\tesla}, thereby tuning the microwave cavity and magnon frequencies to be equal. The excitations of the hybrid system are then two magnon-polariton modes of angular frequencies $\omega_\pm=\omega_c\pm g$. Using the AWG we apply waveforms to the IQ demodulator corresponding to a single tone Floquet drive such that the time-varying angular frequency of the microwave mode is again given by $\omega(t) = \omega_c + a \sin (\Omega t)$. We measure S21 as before.

\begin{figure}
\includegraphics{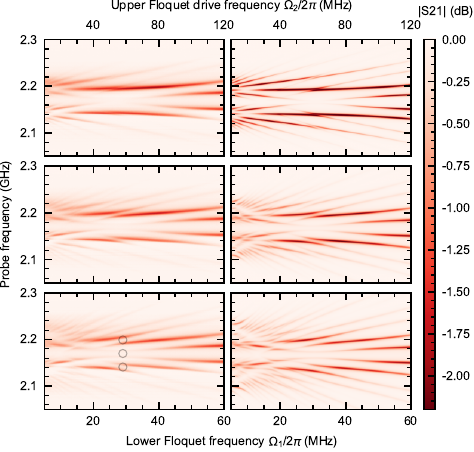}
\caption{\label{fig:f-2f}\Sto of feedline for two-frequency cavity modulation with $\Omega_2 = 2\Omega_1$, $a_1/2\pi=\qty{40}{\mega\hertz}$, and $\phi= 0$. $\Omega_1/2\pi$ is stepped from \qty{5}{\mega\hertz} to \qty{60}{\mega\hertz}. Rows from top to bottom correspond to $a_2/2\pi = \qty{10}{\mega\hertz}$, $a_2/2\pi = \qty{25}{\mega\hertz}$, $a_2/2\pi = \qty{40}{\mega\hertz}$. Left column shows experimental data; right column -- input-output theory (see main text).}
\end{figure}

In Fig.~\ref{fig:ExperimentConfiguration}(e) we show \Sto as a function of Floquet frequency $\Omega/2\pi$ and probe frequency, with the amplitude of the Floquet drive fixed at $a/2\pi = \qty{50}{\mega\hertz}$. Sidebands are seen on both magnon-polariton branches, spaced at integer multiples of $\Omega/2\pi$  from the underlying modes. At the point where $\Omega/2\pi = \qty{41.4}{\mega\hertz}=2g/2\pi$, we see an anticrossing between the first upper (lower) sideband of the lower (upper) magnon-polariton branch ($\omega_\mp \pm \Omega$) and the upper (lower) magnon-polariton branch ($\omega_\pm$). We also see a family of anticrossings between higher order modes at lower Floquet frequencies $\Omega$, following the rule that for finite coupling sizes $\Delta n$ must be odd. For example, the sideband of order $-1$ of the upper branch ($\omega_+ - \Omega$) and the sideband of order $+1$ of the lower branch ($\omega_- + \Omega$) do not anticross with each other.

We next study the dependence of the size of the anticrossing between $\omega_\pm$ and $\omega_\mp \pm \Omega$ on the amplitude of the Floquet drive, $a$. We fix $\Omega/2\pi = \qty{41.4}{\mega\hertz}=2g/2\pi$ (dashed line in Fig.~\ref{fig:ExperimentConfiguration}(e)), which is the low power limit of the position of the anticrossing between these modes. In Fig.~\ref{fig:ExperimentConfiguration}(f) we show \Sto for this value of $\Omega$ as a function of Floquet amplitude $a/2\pi$ and probe frequency. We see that the coupling between sideband and magnon-polaritons increases with increasing $a$.

The quasienergy level spectrum measured so far, while extended to higher drive amplitudes, is consistent with previous measurements on cavity magnon-polariton systems in which the Floquet drive was applied to the magnon population\cite{xuFloquetCavityElectromagnonics2020}. However, the visibility of the modes in S21 measurements is different here to that seen in Ref.~[\!\!\citenum{xuFloquetCavityElectromagnonics2020}]. We ascribe this to the fact that in our case the photon mode, through which the behaviour of the system is probed, is modulated rather than the magnon mode. This leads to destructive interference between the $n-1$ sideband of the upper magnon-polariton branch, and the $n+1$ sideband of the lower magnon-polariton branch, rendering the mode dark.

\begin{figure}[b]
\includegraphics{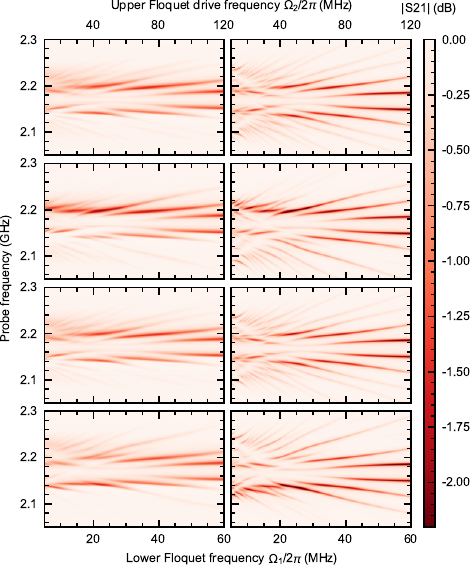}
\caption{\label{fig:f-2f-phase}\Sto of feedline for two-frequency cavity modulation with $\Omega_2 = 2\Omega_1$, $a_1/2\pi=\qty{35}{\mega\hertz}$, and $a_2/2\pi = \qty{40}{\mega\hertz}$. $\Omega_1/2\pi$ is stepped from \qty{5}{\mega\hertz} to \qty{60}{\mega\hertz}. Rows from top to bottom correspond to $\phi = 0$, $\phi = \pi/2$, $\phi = \pi$, $\phi = 3\pi/2$. Left column shows experimental data; right column -- input-output theory.}
\end{figure}

We now move beyond single-tone driving to multifrequency modulation. We apply commensurate two-frequency Floquet drives such that 
$\omega(t) = \omega_c + a_1 \sin (\Omega_1 t) + a_2 \sin (\Omega_2 t + \phi)$, where $\Omega_2 = m \Omega_1$ and $m$ is integer and small. We start by applying waveforms for which $m=2$. In the left column of Fig.~\ref{fig:f-2f} we show \Sto as a function of $\Omega_1$, with $\Omega_2 = 2 \Omega_1$ (shown on the upper $x$-axes), for $a_1/2\pi = \qty{40}{\mega\hertz}$, $\phi=0$, and (panels from top to bottom) $a_2/2\pi=\{\qty{10}{\mega\hertz}$, \qty{25}{\mega\hertz}, \qty{40}{\mega\hertz}\}. The multi-tone driving introduces modifications to the quasimode spectrum; in particular anticrossings are now introduced between sidebands for which $\Delta n$ is even. Most noticeably, anticrossings open between the $\omega_+ - \Omega$ and $\omega_- + \Omega$ sidebands, the $\omega_+$ mode and $\omega_- + 2\Omega$ sideband, and the $\omega_- $ mode and $\omega_+ - 2\Omega$ sideband; these are highlighted with grey circles in the lower left panel of Fig.~\ref{fig:f-2f}. Within this range of second Floquet amplitude $a_2$, the anticrossings dependent on the upper frequency increase in size with increasing $a_2$. Conversely, the size of those anticrossings which do not rely on the presence of the drive at $\Omega_2$, such as those at $\Omega_1\approx\qty{47}{\mega\hertz}$, are not influenced by $a_2$ in its measured range.

\begin{figure}
\includegraphics{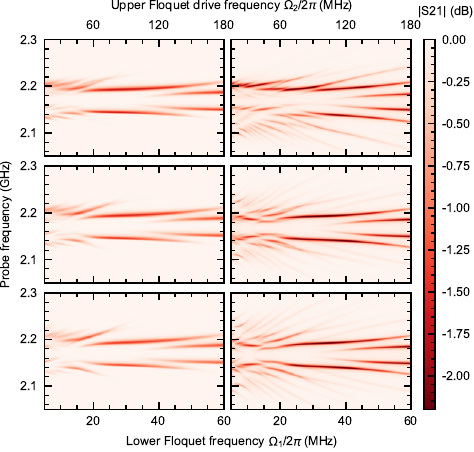}
\caption{\label{fig:f-3f}\Sto of feedline for two-frequency cavity modulation with $\Omega_2 = 3\Omega_1$, $a_1/2\pi=\qty{32.5}{\mega\hertz}$, and $\phi= 0$. $\Omega_1/2\pi$ is stepped from \qty{5}{\mega\hertz} to \qty{60}{\mega\hertz}. Rows from top to bottom correspond to $a_2/2\pi = \qty{20}{\mega\hertz}$, $a_2/2\pi = \qty{30}{\mega\hertz}$, $a_2/2\pi = \qty{40}{\mega\hertz}$. Left column shows experimental data; right column -- input-output theory.}
\end{figure}

The time-varying system can be modeled using the Floquet matrix formalism~\cite{poertnerValidityManymodeFloquet2020, shirleySolutionSchrodingerEquation1965} for the time-periodic Hamiltonian
\begin{align}
H(t) &= H_0 + V_{\mathrm{dr}}(t) \\
\hbar^{-1} H_0 &= \omega_c c^\dagger c + \omega_m m^\dagger m + g\, c^\dagger m + g^* m^\dagger c \\
\hbar^{-1} V_{\mathrm{dr}}(t) &= c^\dagger c\left(a_1 \sin(\Omega_1 t) + a_2 \sin(\Omega_2 t + \phi) \right).
\end{align}
By expanding this in terms of its harmonics we write it in matrix form. The $n\textrm{th}$ $2\times2$ diagonal block is of the form
\begin{align}
    \begin{bmatrix}
    \omega_c +n\Omega_1& g \\
    g & \omega_m +n\Omega_1
    \end{bmatrix}
\end{align}
(where in our experiments $\omega_c=\omega_m$) and describes the underlying magnon-polariton system. In our calculations we truncate the matrix at sidebands of order $n = \pm 8$.
The multi-frequency Floquet coupling is modelled with off-diagonal blocks of the form
\begin{align}
    \begin{bmatrix}
    \pm \frac{1}{2} i a_1 & 0 \\
    0 & 0
    \end{bmatrix}
\end{align}
between diagonal blocks $n$ and $n\pm1$ corresponding to the coupling due to $\Omega_1$, and of form
\begin{align}
    \begin{bmatrix}
    \pm \frac{1}{2} i a_2e^{\mp i\phi} & 0 \\
    0 & 0
    \end{bmatrix}
\end{align}
between blocks $n$ and $n\pm m$ corresponding to the coupling due to $\Omega_2$. Comparison to experimental results is carried out using cavity input-output theory~\cite{gardinerInputOutputDamped1985}, with the input mode coupling to the $n=0$ cavity mode. In the right column
of Fig.~\ref{fig:f-2f} we show \Sto calculated using this model, with the same parameters as the experimental results in the left column, and the loss rate of the magnon modes~\cite{lambertIdentificationSpinWave2015} set to $2\pi\times\qty{5}{\mega\hertz}$. We find excellent agreement between the two.

The inclusion of a second drive tone introduces another degree of freedom in addition to the modulation depth -- the relative phase between the two tones. In Fig.~\ref{fig:f-2f-phase} (left column) we fix $a_1/2\pi=\qty{32.5}{\mega\hertz}$ and $a_2/2\pi = \qty{40}{\mega\hertz}$. The phase $\phi$ is varied within each column in steps of $\pi/2$, starting with $\phi = 0$ in the uppermost panels. At phase differences of $\phi = \pi/2$ and $\phi = 3\pi/2$ we observe asymmetric sidebands, such that the visibilities are different between upper and lower sidebands with positive and negative $n$. This is due to each sideband being generated by multiple paths characterised by Bessel functions of different order $n$; for example a sideband at $\omega_-+2\Omega_1$ has a contribution from the second order sideband due to $\Omega_1$, of strength $J_2(a_1/\Omega_1)$, and a contribution from the first order sideband due to $\Omega_2$, of strength $J_1(a_2/\Omega_2)$. The phase shift between the two drive frequencies results in either positive or negative interference between alternative sideband generation paths involving different signs of $n$, and hence different signs of the corresponding Bessel functions\cite{janikTwotoneFrequencymodulationSpectroscopy1986, ProakisSalehiCSE2}. Our results are again in good agreement with multi-frequency Floquet theory and input-output theory (Fig.~\ref{fig:f-2f-phase}, right column).

Finally, we apply waveforms such that $\Omega_2=3\Omega_1$ with $\Omega_2$ up to \qty{180}{\mega\hertz}, fix $\phi=0$, and again measure S21 (Fig.~\ref{fig:f-3f}). Further structure is introduced to the spectrum. We note in particular, the condition that $\Delta n$ must be odd for finite coupling between quasi-energy levels is restored; for example, level crossings are introduced between $\omega_+ - \Omega$ and $\omega_- + \Omega$. This is due to the coupling Hamiltonian once more having only odd terms when expanded in terms of Bessel functions. Our results are well described by Floquet theory as before (Fig.~\ref{fig:f-3f}, right column).

In conclusion, we have demonstrated Floquet engineering in a cavity magnon-polariton system in which the photon frequency rather than the magnon frequency is modulated. This gives us access to large depth and frequencies of modulation, allowing multifrequency Floquet drives to be applied. We demonstrate that, as well as the drive amplitudes, the relative phase between the drives affects the resulting quasi-energy spectrum, and leads to asymmetric sidebands. Our results are described well using a Floquet Hamiltonian and input-output theory. Our device architecture is a versatile platform for the study of physics in which cavity mode parameters including both frequency and lifetime~\cite{lambertCoherentControlMagnon2025} are varied in time.

\section*{Acknowledgments}

WW and NJL are supported by {\it Quantum Technologies Aotearoa}, a research program funded by the New Zealand Ministry of Business, Innovation and Employment, contract number UOO2347. NJL acknowledges funding from Royal Society of New Zealand Marsden Grant no. 24-UOO-153.

\section*{Author Contributions}

NJL conceptualized and supervised the experiment.
LH performed experimental measurements and data analysis. ARM and WW carried out theoretical analysis and contributed to data analysis. WW and NJL prepared the manuscript.

\bibliography{FloquetMultifrequency}

\end{document}